\documentclass[a4paper,10pt,twoside]{cpc-hepnp}

\usepackage{multicol}
\usepackage{graphicx}
\usepackage{booktabs}
\usepackage{amssymb,bm,mathrsfs,bbm,amscd}
\usepackage[tbtags]{amsmath}
\usepackage{lastpage}
\usepackage{CJK}

\begin{document}
\begin{CJK*}{GBK}{song}

\fancyhead[c]{Submitted to `` Chinese Physics C " }


\title{New-Type Internal Target for Structural Ion Stripping}

\author{%
      D. N. Makarov$^{1,2;1)}$\email{makarovd0608@yandex.ru}%
\quad V. I. Matveev$^{1,2}$
\quad 
\quad 
}

\maketitle

\address{%
$^1$ Federal Center for Integrated Arctic Research, Russian Academy of Sciences, Arkhangelsk 163000, Russia.\\
$^2$ Northern (Arctic) Federal University, Arkhangelsk 163002, Russia.\\
}

\begin{abstract}
The search is now on for, new materials that can be used for ionic stripping. Materials that maximize the stripping of the structural ion are important for conducting experiments with quark-gluon plasma. Although this paper is a theoretical study, it offers practical application, in heavy-ion accelerators, of the new effect of collision multiplicity with high-energy ions interacting with polyatomic targets. It is shown that internal nanostructured targets in which the collision multiplicity effect is manifested can more efficiently strip out structural ions compared to standard internal targets for stripping. A target consisting of oriented nano-tubes with the $C_{240}$ chirality (10,0) is considered as an example. A comparison with the stripping process on a carbon target with the same number of misaligned atoms in a unit of volume $C$ is provided.
\end{abstract}

\begin{keyword}
Ion stripping, Accelerator, Target, Ion beam, Nanotube, Charge composition.
\end{keyword}

\begin{pacs}
29.25.-t; 29.27.-a; 34.70.+e
\end{pacs}


\begin{multicols}{2}

\vspace{6mm}

\section{Introduction}
The current surge of interest in high-energy collisions involving highly charged ions is associated with the development and application of modern heavy-ion accelerators. The effect of collision multiplicity - which contributes significantly to the cross-sections of non-elastic processes at the relative collision velocities that greatly exceed the characteristic atomic velocity - may relate to new and interesting effects that follow high-energy ion collisions. The effect was first predicted in theoretical studies by \cite{Matveev08, MGM2009, MMR2010}; the most convincing arguments in pure physics that are used to study the effect of collision multiplicity are based upon a consideration of collisions with nanotubes oriented along the ion velocity. It should be stressed that there is still no experimental proof of the effect; such proof would be technically difficult and would require a heavy-ion accelerator. However, the effect of these collisions is of interest not only because of the fundamental physics involved, but also because potentially they have considerable practical application. Hence in this paper we put forward a model of a nanoscale structure of internal targets in accelerators, that could not only increase the degree of ion-beam stripping, but also, serve as the direct experimental proof of an exceptionally interesting effect of collision multiplicity. Moreover, from a theoretical standpoint at least, we consider that our simple qualitative explanation provides a strong basis for the existence of such an effect.

In general, solid targets are used in most experiments associated with fast-ion beam stripping. The choice of such targets is fully justified because the stripping process is more efficient in them than in gas targets \cite{Ryding1970, Ogawa2007,Art2008}. To improve stripping, the gas density of the gas targets must be increased. In contrast, when working with solid targets, one is able to simply choose the target with the best properties for stripping. There is a current requirement for the maximum stripping of a structured heavy ion for experiments with quark-gluon plasma and for other heavy-ion accelerators, for example, for the NICA Project (Dubna, Russia) \cite{Agapov2016}, RHIC (USA), GSI (Darmstadt, Germany) and LHC (Switzerland, France). Structured heavy-ion stripping is more significant at $v>>1$, where  $v$ is the ion velocity in atomic units (used from here on)). Even at the ultra-relativistic velocities of a heavy ion its stripping cannot be complete. Various semiempirical expressions are often used for the assessment of structured ion stripping down to some average (equilibrium) charge  $Z=\sum_{i} i F_{i} $, where $F_{i}$ is relative amount of ions with the charge $i$ in the equilibrium distribution. For example, the formula obtained by Northclife \cite{Northcliffe}, is often used: 
\begin{equation}
Z=Z_{0}\left(1-e^{-v/{Z^{2/3}_{0}} }  \right) , 
\label{1St}
\end{equation}
where $Z_{0}$ is the charge of a completely stripped ion. Expression  (\ref{1St}) is often modified by the introduction of various parameters determined on the basis of best agreement between theoretical and experimental approaches. We should mention that modifying expression  (\ref{1St}) does not change it much, but it does improve its accuracy expressed by several percents of the experiment. Apart from expression  (\ref{1St}) and its modifications other expressions are used, also obtained semiempirically \cite{Dmitriev1965, Shima1984}.

This paper shows that the charge composition of an ion beam can be significantly changed if targets showing the effect of collision multiplicity are used, since this effect increases  \cite{MGM2009, MMR2010}the ionization cross-sections due to multiple collisions. In fact, the equilibrium charge composition in the ion beam is determined by the ionization and recharge cross-sections \cite{Nikolaev1965, Betz1972}. If we select a material that can significantly increase the ionization cross-section at the unchanged recharge cross-section, this will result in significantly better stripping. How collision multiplicity can help to significantly improve the ion-beam stripping can be qualitatively explained  \cite{Matveev08, MGM2009, MMR2010}. The easiest way to illustrate the contribution of collision multiplicity is to consider the example of the collision of highly charged structured ions with a two-atom molecule (consisting of two identical atoms). The ion moving at a relativistic velocity travels the distance between the atoms of the molecule in a time expressed as $\tau _{c}$ which is of the order of or less $10^{-19}$ seconds. It is evident, that this time is much shorter than the mean excited-state lifetime $\tau_{r}$ of the ion (projectile) relative to radiational and Auger decays; therefore, these processes at the sequential collision of the structured ion with the atoms of the molecules cannot be taken into account. Let us assume that  $\tau _{c}$ is so small that we can neglect the evolution of the electron states of the ion (i.e. $\tau_{c}<<\tau_{e}$, where $\tau_{e}$ is the characteristic time for the ion electrons) during the intervals between two sequential collisions. Let the projectile electrons, at the collision with two centers of the molecule, acquire an momentum ${\bf q}_1$ from the first center and the momentum ${\bf q}_2$ from the second center; total momentum then acquired by the ion electrons is ${\bf q}={\bf q}_1+{\bf q}_2$. 
When the molecule axis is horizontal relative to the ion velocity, the projectile collides with two atoms of the molecule and ${\bf q}_1={\bf q}_2$ and therefore ${\bf q}^2={\bf q}_{||}^2=(2{\bf q}_1)^2$. When the molecule axis is perpendicular, the ion moving along a straight trajectory collides either with one atom of the molecule, or with the other atom, and then either ${\bf q}_1\not=0$ and ${\bf q}_2=0$, or ${\bf q}_1=0$ and ${\bf q}_2\not=0$. 
Thus, at the parallel orientation, the momentum acquired by the ion electrons is twice as large as that for a perpendicular orientation. The probability of the ion ionization is proportional to the squared aggregate momentum ${\bf q}^2=({\bf q}_1+{\bf q}_2)^2$, acquired at the collision, so that the probability of the ion electron shells at the parallel orientation of the molecule is four times as large as the ionization probability at the perpendicular orientation. It is clear that similar arguments apply to the collisions of sufficiently fast structured ions with the molecules consisting of more than two atoms, or with more complicated targets (e.g. with nanotubes). Strictly speaking, during the calculations of the ionization processes the total (not only as  ${\bf q}^2$) dependence of nonelastic form factors of the ion electrons on the acquired momentum should be taken into account and integrated over the impact parameters. Evidently, this alters the presented qualitative estimations, but the effect of the change in ion ionization cross-section change at the change of the target orientation remains significant and can result \cite{Matveev08, MMR2010} in the increase of the projectile ionization cross-section by several times. It should be mentioned that in this paper we will consider the equilibrium distribution of the charge content of the ion beam after the latter has passed through the target.

\vspace{6mm}

\section{ Main part}

For the calculation of the cross-section of the relativistic ion ionization by the field of a neutral multi-electron atom let us use the eikonal formula (13) from \cite{Viotkiv2002} (also see  \cite{Viotkiv}).  The amplitude of the probability of the ion electron transition from the state $| \phi_{0}>$ with the energy  $\varepsilon_{0}$ to the state  $|\phi_{n}>$ with the energy $\varepsilon_{n}$ is \cite{Viotkiv2002}  
\begin{eqnarray}
a_{n0}=<\phi_{n}\mid (1-\alpha_{z})e^{i(\varepsilon_{n}-\varepsilon_{0})z/c}\times
\nonumber\\
\times \exp\biggl(-i\frac{2Z_{a}}{c}\sum\limits_{i = 1}^{3}A_{i}K_{0}(\kappa_{ i}|{\bf b}-{\bf r}_{\bot}|)
\biggr)\mid \phi_{0}>,
\label{11aa}
\end{eqnarray}
where $K_{\nu}(z)$ is the Macdonald function, ${\bf b}$ is the impact parameter,
${\bf r}_{\bot}$ is the ion electron coordinate projection on the plane of the impact parameter, the ion velocity is along the $z$ axis,
$\alpha_{z}$ is the $z$ - component of the Dirac matrix $\mbox{\boldmath $\alpha$}=(\alpha_x, \alpha_y,\alpha_z)$.
In the formula (\ref{11aa}) the eikonal phase is calculated in the Dirac-Hartree-Fock-Slater model \cite{Salvat} according to which the potential created by the stationary neutral atom at the origin of coordinates is
\begin{equation}
\varphi'({\bf r}) = \frac{Z_{a}}{|{\bf r}|}\sum\limits_{i = 1}^{3}A_{i}\exp\left(-\kappa_{ i}|{\bf r}|\right)\:,
\label{forb}
\end{equation}
where $Z_{a}$ is the atom nuclei charge, $A_{i}$ and $\kappa_{i}$ are the constants tabulated in \cite{Salvat} for all atom elements with $Z_{a} = 1,2,\ldots,92$, ${\bf r}$ are the coordinates of the observation point. Note, that the small $Z_{a}/c$  correspond to the applicability of the perturbation theory, and it is easy to see that in this case (\ref{11aa}) coincides with the ultrarelativistic limit of the known formula for the amplitude in the first order of the perturbation theory over the atom field given in \cite{Viotkiv}. 
However, if we consider the fast ion collision with a polyatomic system, than the contribution to the eikonal phase in the expression (\ref{11aa}) can be not small even at small $Z_{a}/c$ and the use of the perturbation theory will be incorrect. In the formula (\ref{11aa}) the field of application $r_{\bot}$ is limited by the transversal size of the heigh charge ion that is significantly smaller than one, while the transversal size of the neutral atom is of the order of unity. Therefore the mean atom field can be considered to be homogeneous on the ion sizes, which corresponds the decomposition into small  $r_{\bot}/b$ using the formula
\begin{equation}
K_{0}(\kappa|{\bf b}-{\bf r}_{\bot}|)\approx K_{0}(\kappa b)+K_{1}(\kappa b)
\frac{\kappa{\bf b}{\bf r}_{\bot}}{b}.
\label{k0}
\end{equation}
The term $K_{0}(\kappa b)$, as the one that does not causes electron transitions, can be omitted, and in the result the formula (\ref{11aa}) at the orthogonal  $|0>$ and $|n>$ will have the form 
\begin{eqnarray}
a_{n0}=<\phi_{n}\mid (1-\alpha_{z})e^{i(\varepsilon_{n}-\varepsilon_{0})z/c}
e^{-i{\bf q}{\bf r}},
\biggr)\mid \phi_{0}>,
\label{11aaa}
\end{eqnarray}
where the vector ${\bf q}$ has the sense of an momentum acquired by the ion electron at its collision with atoms and is
\begin{equation}
{\bf q}=\frac{2Z_{a}}{v}\sum\limits_{i=1}^{3}
\kappa_{i}A_{i}K_{1}(\kappa_{i}b)\frac{{\bf b}}{b}\;.
\label{q_0}
\end{equation}
In the case of the collision with a polyatomic target, the potential $\varphi$ is the sum of the potentials from individual atoms included in the target:
\begin{equation}
\varphi= \sum\limits_{m = 1}^{N}\varphi_{m}\:,
\label{for3}
\end{equation}
where $\varphi_{m}$ is the potential created by the $m$-th atom of the target; $m=1,2,...,N$, here $N$ is the number of atoms in the target. In the rest frame of the ion (\ref{for3}) the potential turns out to be dependent on the time and impacts the ion electrons during some period $\sqrt{1-v^2 /c^2}L/c$ (where $L$ is the characteristic size of the target). Let us assume that this time is significantly shorter than the characteristic periods of time for the ion electrons. Than the ion electrons percept the collisions with the target atoms as instantaneous and simultaneous ones. Accordingly, in (\ref{11aaa}) the vector ${\bf q}$ is the sum of the momenta transmitted by the ion electrons at the collision with the atoms of the target with $N$ atoms, and is
\begin{equation}
{\bf q}=\sum\limits_{m = 1}^{N}{\bf q}_m=\frac{2Z_{a}}{v}\sum\limits_{m = 1}^{N}\sum\limits_{i=1}^{3}
\kappa_{i}A_{ i}K_{1}(\kappa_{i}b_m)\frac{{\bf b}_m}{b_m}\;,
\label{q_m}
\end{equation}
where ${\bf b}_m$ is the impact parameter relative to the $m$-th atom, it is evident that if the target geometry is fixed, that all ${\bf b}_m$ and unambiguously connected with a impact parameter ${\bf b}$calculated from some atom of the target. Thus, at the ion collision with the $N$-atom target, the ion electron excitation amplitude shall be calculated over the formula (\ref{11aaa}), where ${\bf q}$ is expressed by the formula (\ref{q_m}). At $N=1$ in (\ref{q_m}) the formula (\ref{11aaa}) describes a collision with a one-atom target. In all cases the corresponding cross-section is calculated over the formula
\begin{eqnarray}
\sigma_{n0} = \int|a_{n0}|^{2}d^{2}{\bf b}\;.
\label{8mh1}
\end{eqnarray}

For the calculation of the recharge cross-sections we will take into account the dominating channel only in the case of high-charge ions moving at a relativistic velocity, and use the Stobbe formula (see, e.g., \cite{Scheidenberger1998}) for the radiation recharge that describes these cross-sections for the relativistic ions quite well \cite{Scheidenberger1998}
\begin{equation}
\sigma = 3.273 \times 10^{-4} Z_{t} \left(\frac{\xi^3}{1+\xi^2} \right)^2 \frac{e^{-4\xi \times  \arctan(\xi^{-1})}}{1-e^{-2\pi \xi } },
\label{Stobbe}
\end{equation}
where $\xi= 1/\sqrt{\eta}$, $\eta = E_{kin}/E_{K}$, $E_{kin}$ is the kinetic energy of the target electron in the ion rest system; $E_{K}$ is the energy of the electron bond on the ion  $K$-shell, $Z_{t}$ is the amount of the target ions.

Let us choose the target where this effect shows itself to the maximum. This can be a chain of atoms located along the ion velocity vector - an oriented target. The real system where such chains can be found, can be, e.g., a nanotube. For the sake of argument, let us consider a nanotube $C_{240}$ with the chirality (10,0), containing 20 of such chains parallel to the nanotube main axis, at that each chain includes $N=12$ atoms of carbons. Then let us consider the structured ion ionization cross-section at the same nano-tube, with the assumption that the main axis of the nanotube is parallel to the ion velocity vector. Let us consider the high-charge structured ions with the visible charge $Z_P$ significantly more than unit (e.g., for the gold ion $Au^{76+}$ charge $Z_P=76$). Then, in the model of the neutral atoms that comprise the target the amplitude and the cross-section of the projectile electron transition from the sate $|0>$ to some other state $|n>$  can be found over the formulas (\ref{11aaa}) and (\ref{8mh1}). As we deem that the nanotube is strictly built-inn along the ion velocity , and since the size of the ion electron coat is much less than the characteristic atomic size, we can assume that the ion interacts with one chain of atoms only. There are 20 of such chains, so the calculation for $Z_P\gg 1$ shall be performed as follows. First, the cross-section for the collision with one chain shall be found, than this result shall be multiplied by 20 - the number of chains in the nanotubes. At that, within one chain the momenta acquired at the collision with each atom, are the same and are, e.g., ${\bf q}_1 $, so that in (\ref{q_m}) 
${\bf q}=\sum\limits_{m = 1}^{N}{\bf q}_m =N {\bf q}_1 $. It is easy to see that this is the momentum ${\bf q}=N {\bf q}_1 $ with which the amplitude (\ref{11aaa}) of non-elastic ion electron transitions at the collision with a chain of atoms shall be calculated, at that the calculation over the formula (\ref{8mh1}) we shall assume that in the formula (\ref{q_m}) all impact parameters ${\bf b}_m$ equal to each other and in (\ref{8mh1}) ${\bf b}={\bf b}_1$ . As for the recharge cross-section on such nanotube, its value is obtained by the multiplication of the recharge cross-section on one atom by the number of atoms in the nanotube.

For the calculations of the equilibrium charge composition distribution in the ion beam $F_{i}$ we use the known equations and approaches \cite{Nikolaev1965, Betz1972}:
\begin{equation}
\sum_{j\leq i, k >i}F_{j}\sigma_{j,k}=\sum_{j>i, k\leq i}F_{j}\sigma_{j,k}~;~~  \sum_{i}F_{i}=1 ,
\label{FNik}
\end{equation}
where $\sigma_{j,k}$ is the process cross-section at which the ion with the charge $j$ turns into the ion with the charge $k$, and $i=0,1,2,...,Z_{0}$.
In the cases when we can neglect the loss and the capture of two and more electrons (as the cross-sections of these processes are much smaller that the one-electron cross-sections), the first formula of the expressions (\ref{FNik}) gains the form  $F_{i}\sigma_{i,i+1}=F_{i+1}\sigma_{i+1,i}$. By solving the system of equations (\ref{FNik}) with account for the one-electron losses and the captures, we find $F_{i}$. In these equations we nee to know only the ionization and recharge cross-sections. It is difficult to determine in the considered case which energy states the ion beam ionization processes are initiated from, and which states the capture processes go to. Therefore we assume, that the radiation capture processes go to the K-shell of the ion, and the ionization processes go not from the excited states of the ion, but from the main ones. We have checked that such supposition for standard targets ($Cu$ - copper,  $Al$ - aluminium, $Au$ - gold), agrees well with other papers \cite{Shevelko_UFN_2013} and the experiment \cite{Scheidenberger1998}. The calculation can be simplified if we calculate only those $F_{i}$ that significantly differ from zero, and all the rest ones will be considered as zero (in our calculation it is $F_{i}<10^{-5}$). It is rather easy to evaluate in such case, e.g., for $Au, Pb, U$, that for the amount of the ion electrons that participate in the stationary regime in the ionization and recharge processes the number of relativistic ion velocities will be not more than three. E.g., for  $Au$ we have that $F_{79},F_{78},F_{77}$ have not small values, but $F_{76}<10^{-5}$, so all the rest $F_{75},F_{74}, ..., F_{1}$ can be not taken into account. 

\vspace{6mm}
\section{Calculation of charge states of an ion beam}

Let us show the calculations of the equilibrium charge states of an ion beam for two targets: 1) the target consisting of nanotubes parallel to the projectile velocity $C_{240}$  with the chirality (10.0) and the length of the chains along the main axis  $L=47$ at.units. 2) the target consisting of chaotically located isolated atoms of carbon. In both cases the average densities of the carbon atoms in the targets are the same so that there are 240 atoms in the nanotube volume, for the first and the second targets. We choose the following projectiles for these targets:  $Au$ - gold, $Pb$ -  lead, $U$ uranium. The energies of the projectiles shall be chosen so that the condition  $\sqrt {1-v^2/c^2}L /v\ll \tau_e$ is fulfilled on one nanotube, and more over - on an isolated atom. Therefore we choose the kinetic energy suitable for all considered projectiles $E=1TeV/n$.The ion velocity shall be considered to be unchanged. The calculation results that not zero (to more exact, $F_{i}>10^{-5}$) $F_{i}$ are given in Tables 1-3 for the projectiles  $Au$, $Pb$, $U$, respectively, and for two targets - nanotube  $C_{240}$  - the middle line, and for carbon - the lower line. 
To compare and evaluate the effect studied, the tables show single-ionization cross sections (in atomic units): the hydrogen-like ion - $ \sigma_ {H} $, the helium-like ion - $ \sigma_ {He} $, the lithium-like ion $ \sigma_ {Li} $. It can be seen from the tables that the effect studied significantly increases the ionization cross sections, which leads to an increase in the stripping of the ion beam.

Now let us estimate the angle of the nanotube orientation within which the effect of collision multiplicity takes place. The calculation results $F_i$ shown in the Tables in the middle lines will be designated as  $F_i^{max}$, and in the lower lines - as $F_i^{min}$. The neutral atom field disappears beyond the atom size, so, to collide with the atom, the ion needs to get to the spot with the size of the order of the transversal size of the atom and the radius of the order of unity. We take into account that the size of the electron coat of a high-charge projectile is much less than the size of the neutral atom as the target component. Assume, that the nanotube length along the mane axis is  $L$, than the projectile that collides with the first atom at the end of the tube collides the last one (the twentieth atom along the main axis), if its velocity direction belongs to the angle interval $0<\theta< \theta_0 \sim 1/L$.   It is evident, that the stripping process depends on the density of the atom electrons that changes exponentially with the distance form the atom nuclei. Thus, it is reasonable that the quantities $F_i$ 
change exponentially from $F_i^{max}$ to $F_i^{min}$ with the change of the nanotube axis orientation angle $0<\theta< \theta_0 $: 
\begin{equation}
F_i(\theta)=F_i^{min}+(F_i^{max}-F_i^{min})e^{-\theta/\theta_0}.
\label{Fk}
\end{equation}
At that the effect of the collision multiplicity reaches its maximum at  $\theta=0 $ and disappears at $\theta>\theta_0 $. 
Therefore, no strict direction of the nanotubes along the ion beam velocity vector is not required for the effect of the collision multiplicity, which means that it is not necessary to direct the ion beam strictly along the nanotube axis. This property of the effect of collision multiplicity enables the experiments on the realized ion beams with the angular dispersion much less than the angle of orientation.
\end{multicols}

\begin{table}[]
\begin{center}
\begin{tabular}{|c|c|c|c|c|c|c|c|}

\hline 

Au            &$F_{76}$&$F_{77}$&$F_{78}$&$F_{79}$&$\sigma_{Li}$&$\sigma_{He}$&$\sigma_{H}$ \\
               &&&&& $10^{-4}$a.u. &$10^{-4}$a.u. &$10^{-4}$a.u.\\
\hline                                                  
$C_{240}$     &0         &0.0004               &0.0274    &0.9722    &1066  &479  &316     \\
\hline
$240 C$           &0.0012         &0.0230               &0.1800   &0.7959    &157  &69  &40   \\
\hline
\end{tabular}
\caption
{
The values of the function of charged states $F_{i}$ and ionization cross sections for the gold ion beams $Au $.}
\label{table:1}
\end{center}
\begin{center}
\begin{tabular}{|c|c|c|c|c|c|c|c|}

\hline 

Pb            &$F_{79}$&$F_{80}$&$F_{81}$&$F_{82}$&$\sigma_{Li}$&$\sigma_{He}$&$\sigma_{H}$ \\
               &&&&& $10^{-4}$a.u. &$10^{-4}$a.u. &$10^{-4}$a.u.\\
\hline                                                  
$C_{240}$     &0         &0.0006               &0.0330    &0.9663    &1000  &449  &296     \\
\hline
$240 C$           &0.0021         &0.0326               &0.2086   &0.7567    &146  &62  &37     \\
\hline
\end{tabular}
\caption
{The values of the function of charged states $F_{i}$ and ionization cross sections for the plumbum ion beams $Pb $.}
\label{table:2}
\end{center}
\begin{center}
\begin{tabular}{|c|c|c|c|c|c|c|c|}

\hline 

U            &$F_{89}$&$F_{90}$&$F_{91}$&$F_{92}$&$\sigma_{Li}$&$\sigma_{He}$&$\sigma_{H}$ \\
               &&&&& $10^{-4}$a.u. &$10^{-4}$a.u. &$10^{-4}$a.u.\\
\hline                                                  
$C_{240}$     &0.00003         &0.0019               &0.0579    &0.9402    &818  &367  &241     \\
\hline
$240 C$           &0.0103        &0.0869               &0.3025   &0.6003    &118  &50  &30     \\
\hline
\end{tabular}
\caption
{The values of the function of charged states $F_{i}$ and ionization cross sections for the uranium ion beams $U$.}
\label{table:3}
\end{center}
\end{table}
\begin{multicols}{2}

\section{Discussion and conclusion}
With that said we show below that new targets (not yet realized) could be created for stripping structured heavy ions, and oriented nanotubes along the ion velocity vector could serve as such targets. Structurally, such targets could be realized in the form of nanotubes parallel to each other, i.e. a ``nanotube forest" consisting of nanotubes stacked as ``logs of firewood", and more complex forms comprising several parallel ``logs" of nanotubes stacked like slices of bread. We should add that it is not necessary to adjust the ends of the nanotubes so that they line up with each other: in principle, it is sufficient to arbitrarily ``stack up" the target out of nanotubes with the axis orientation within the above-described angle $\theta_0 \sim 1/L$. 
It should be mentioned, that the charge composition of the ion beam is determined not by the ion beam interacting with one nanotube, but by it interacting with a macroscopic system comprising such nanotubes. The sizes of such a system would be determined by the fact that the charge composition of the ion beam of such a macroscopic system would be that determined at equilibrium. Such an ``equilibrium", thickness of the target  as measured in micrometers has already been achieved. Therefore, such targets can be used in the same manner as standard targets, that is, having a target thickness of tens to hundreds of micrometers, with the diamter being limited to the target station diameter. As mentioned above, the effect of the collision multiplicity is revealed for high-energy collisions only, while it is utterly unreal to accelerate low-charge ions up to high velocities. Therefore, for example, on the Large Hadron Collider, the process of complete lead-ion stripping takes place in the region of the second target station where the preliminarily stripped lead  $Pb^{(54+)} $ ion beams reach an energy of  $E=5,9 GeV/n $ \cite{Chanel}. This is the area where the target with the proposed design could be located, although such targets have not been developed yet. We would also like to mention, that in our calculation we have chosen the ion energy $E=1TeV/n$ to meet the condition $\sqrt {1-v^2/c^2}L /v\ll \tau_e$ for all electron ion shells (including internal shells) with high charge nuclei. For example, for internal electrons of the uranium ion $ \tau_e \sim 10^{-3} a.u.$, but for $ E = 1TeV/n $, the collision time with the nanotube $ \tau \sim \sqrt{1-v^2 / c^2} L/v \sim 10^{-4} a.u.$. Such energies, and the targets that we have chosen in the form of nanotubes, allow us to determine the effect of collision multiplicity in the, 'pure form', although the effect of collision multiplicity would reveal itself at lower energies, as well for the higher-energy shells where the amount of electrons for heavy ions is much greater than for the internal shells. Therefore, it is absolutely clear at a qualitative level that the effect of collision multiplicity would be significant for the stripping $Pb^{(54+)} $ with the energy $E=5,9 GeV/n $. We note that we have also carried out calculations for the energy $E=5,9 GeV/n $. At this energy, the charge composition of the ion beam did not differ much from the charge composition at an energy $ E = 1TeV/n $, although at $E=5,9 GeV/n $ the evolution of the wave function for the inner electron shells of the ion makes a small contribution, which in combination with the collision multiplicity effect is extremely difficult to calculate. For this reason we have presented a calculation for $ E = 1TeV/n $, in which only the collision multiplicity effect is present, where the processes associated with the evolution of the wave function of electrons in the ion, when colliding with an entire nanosystem, can be neglected. 
However, at the quantitative level, a more detailed description of the stripping process which accounts for the effect of multiplicity is needed, which is extremely difficult due to the complexity of the considered processes. Strictly speaking, we must determine the cross-sections of loss and ion-electron capture for each excited ion state, but the issue of the determination of these excited states remains open. Therefore, for such an ion energy, where it is impossible to neglect the evolution of the electron wave function, it is necessary to develop a theory. Obviously, such a theory must be nonperturbative. At present, such processes can only be calculated numerically for a limited number of target atoms. It is evident that, apart from the nanotubes, there could be other targets where the effect of collision multiplicity reveals itself, for example, conventional monocrystals (e.g. $Cu$) can serve as such targets, but the axes of the lattice in such targets should be oriented over the ion-beam velocity vector. We should mention that the calculation of the stripping which takes account of the effect of collision multiplicity for monocrystals is a difficult issue. These difficulties are associated with the fact that in the monocrystal too many atoms can be built along one line (as long as $L$), which will result in the violation of the condition  $\sqrt {1-v^2/c^2}L /v\ll \tau_e$ at large $L$, even within one chain, and require the electron evolution in the ion to be taken into account (which is extremely difficult). However, in this case, the effect of collision multiplicity would undoubtedly be apparent, and the stripping in monocrystals could dominate even over the stripping in nanotubes, but the calculations would become considerably more complex. It should be added that we consider the effect of stripping an ion beam is determined by the effect of the multiplicity of the collision, which manifests itself through the dimensions of the target where $\theta_ {0} \sim 1 / L$. Therefore, if we choose a single crystal with $\theta_ {0} << 1/L$, then the ion will move along the channel (so-called channeling), and the effect will change and require additional study. In order for the effect to be effective in monocrystals, the ion beam must be oriented so that $\theta_ {0} >> \psi$ ($\psi$ is the channeling angle), but not more than $\theta_{0} \sim 1/L$. The channeling angle is usually very small for heavy ions, so the effect considered by us could also be used in single crystals, but for $\theta_ {0} >> \psi$. It should be added that there have been many studies on the passage of ions in channels, but the effect examined by us has not been observed experimentally, although technicaly they should be observable.

Notwithstanding the approximations in the calculations, and even at the qualitative level, the effect of significant intensification of stripping on new targets is evident. The considered effect for ion stripping is new and has not yet been studied experimentally. In addition, targets for ionic stripping need not be restricted to nanotubes - they could also be single crystals.
\\

\acknowledgments{We thank the Corresponding Member of the Russian Academy of Sciences Meshkova I. N. for discussion and valuable comments.}

\end{multicols}

\clearpage

\end{CJK*}

\end{document}